\documentclass{ws-p8-50x6-00}

\begin{document}

\title{Theory of color confinement: state of the art.}
\author{A. Di Giacomo and G. Paffuti}
\address{Dip. Fisica Universit\`a and I.N.F.N., Sezione di Pisa,\\
Via
Buonarroti 2, Pisa}
\maketitle
\abstracts{The existing evidence for dual superconductivity
as mechanism of color confinement is reviewed. We also discuss
what is known on the dual excitations, which produce confinement by
condensation, and what are the open problems.
}
\section{Established results: confinement and simmetry.}
An almost general consensus exists on the idea that deconfinement is an
order disorder transition\cite{1}. The present experimental upper limits
on the observation of free quarks are $15$ orders of magnitude
smaller than what is expected in the absence of confinement. The only
natural interpretation is that confinement is an absolute property
relying on symmetry.

The confined phase of QCD is disordered (strong coupling). The symmetry
of a disordered phase can be understood in terms of duality\cite{2,3}. The
system should admit non local topological excitations, and, besides the
usual description in terms of local fields a complementary
dual description exists
 in which the
topological excitations become local and their v.e.v. are the order
parameters. The dual effective coupling constant $g_D$ is related to $g$
as $g_D\sim 1/g$. The strong coupling regime is mapped into the weak
coupling of the dual description.

If this is correct, the ultimate goal is to identify the dual excitations
which condense in the confined phase, or at least to identify their
quantum numbers, i.e. the symmetry of the confining vacuum.

On the basis of the above argument we shall disregard all the approaches 
to
the problem which are not based on symmetry. Two possibilities are being
considered for the dual symmetry, which in principle are not mutually
exclusive
\begin{itemize}
\item[1)] Dual excitations carry magnetic charge, i.e. they are
monopoles: their condensation produces dual superconductivity of the
vacuum. Abrikosov electric flux tubes between colored particles are at the
origin of confinement.\cite{4,5,6}
\item[2)] Dual excitations are $Z_N$ vortices, and their condensation is
qualitatively described as ``spaghetti vacuum''\cite{1}.
\end{itemize}
We remark that for many years both ideas have been analyzed with a
zoologist's attitude. Investigations were a more or less ingenious
counting of monopoles and vortices in connection with the deconfining
transition. If useful in the pioneering stages to establish the existence
of the excitations, such kinds of counting give no information on
symmetry.

In a sense also the concept of dominance is similar. Vortices\cite{7} or
maximal abelian monopoles\cite{8} are assumed to be the relevant dual
excitations, their contribution to observables is extracted, one way or
the other, from numerically generated configurations, and shown to give a
good approximation to the full determination. Dominance is  a
necessary condition, but in principle not sufficient to identify the dual
excitations, without further theoretical input. Symmetry is an exact
property and has to be investigated as such.
\section{Monopoles.} A conserved magnetic charge can be associated to any
operator $\vec \varphi(x)$ in the adjoint representation: in what follows
we shall consider $SU(2)$ gauge group for the sake of simplicity.
Extension to higher groups will only be sketched.

Define $\hat\varphi = \vec\varphi(x)/|\vec\varphi(x)|$, a direction in
color space. $\hat\varphi$ is defined everywhere except at zeros of
$\vec\varphi$, and\cite{7}
\begin{equation}
F_{\mu\nu} = \hat\varphi\cdot \vec G_{\mu\nu} - \frac{1}{g}
\hat\varphi\cdot(D_\mu\hat\varphi \wedge D_\nu \hat\varphi)
\label{eq1}
\end{equation}
with $\vec G_{\mu\nu} = \partial_\mu \vec A_\nu - 
\partial_\nu \vec A_\mu + g \vec A_\mu\wedge \vec A_\nu$ the field
strength tensor, and $D_\mu = \partial_\mu - i g \vec A_\mu \wedge$ the
covariant derivative.

The two terms in eq.(\ref{eq1}) are separately gauge invariant and color
singlets: the specific combination is chosen in such a way that bilinears
in
$A_\mu A_\nu$ and $A_\mu\partial_\nu\varphi$ cancel, so that one has
identically
\begin{equation}
F_{\mu\nu} = \partial_\mu(\hat\varphi \vec A_\nu) -
\partial_\nu(\hat\varphi \vec A_\mu) - \frac{1}{g}\hat\varphi(\partial_\mu
\hat\varphi \wedge \partial_\nu\hat\varphi)\label{eq2}
\end{equation}
A gauge transformation bringing $\hat\varphi$ to a fixed direction, say
3 axis (abelian projection) eliminates the second term of eq.(\ref{eq2})
and makes $F_{\mu\nu}$ an abelian field\cite{6}
\[ F_{\mu\nu} = \partial_\mu A^3_\nu - \partial_\nu A^3_\mu\]
The dual field $F^*_{\alpha\beta} = \frac{1}{2}
\varepsilon_{\alpha\beta\mu\nu} F^{\mu\nu}$ defines a magnetic current
\begin{equation}
j_\alpha = \partial^\beta F^*_{\alpha\beta}\label{eq3}
\end{equation}
which is identically conserved $\partial^\alpha j_\alpha = 0$.
In a non compact formulation of the theory $j_\alpha$ of eq.(\ref{eq3}) is
identically zero (Bianchi identity): in a compact formulation it can
be different from zero.

After abelian projection $U(1)$ Dirac monopoles coupled to $F_{\mu\nu}$
show up at zeros of $\vec\varphi$.

For generic $SU(N)$ gauge group cancellation of bilinears identifies a
set of $N-1$ directions in the algebra, $\Phi_i = \varphi_i^a T^a$, 
and each
of them identifies a symmetric subspace of the algebra. The field $\Phi$
can then be written as a superposition of $\Phi_i$
\[ \Phi = \sum_1^{N-1} c_i(x) \Phi_i\]
Monopoles appear at the zeros of $c_i(x)$, as magnetic $U(1)$ charges.

Dual superconductivity will occour when any of the above magnetic charges
condense in the vacuum, so that magnetic $U(1)$ is broken \`a la Higgs.

A disorder parameter (order parameter of the dual system) $\langle \mu
\rangle$ can be defined, which is the vacuum expectation value of an
operator $\mu$ carrying magnetic charge\cite{7,8}.

The operator $\mu$ is magnetic $U(1)$ charged (and gauge invariant).
It is non
local but 
localized enough to
obey cluster property.

$\langle\mu\rangle$ is measured around the deconfining phase transition
on lattices of different spatial size, and then the thermodynamic limit
$V\to \infty$ is performed by use of finite size scaling
techniques\cite{10}.

The result is
\begin{eqnarray}
\langle \mu\rangle \neq 0 && T < T_C\nonumber\\
\langle \mu\rangle = 0 && T > T_C\nonumber\\
\langle \mu\rangle \mathop\simeq_{T\to T_C} \tau^\delta, &&
\tau \equiv 1-\frac{T}{T_C}\label{eq4}
\end{eqnarray}
The other important result is that\cite{11,13}
the behaviour of $\langle \mu\rangle$ is independent of the
abelian projection used to define the magnetic charge. 

In fact the
computation of $\langle\mu\rangle$ is the computation of a partition
function, which is very noisy. All the relevant information is,
however, contained
in the quantity
\begin{equation}
\rho = \frac{d}{d\beta}\ln\langle \mu\rangle\label{eq5}
\end{equation}
$\langle \mu\rangle$ in dimensionless, and depends on the ratios of the
lengths involved, the lattice spacing $a$, the spatial lattice size $L$
and the correlation length $\Lambda$. The time extension of the lattice
identifies the temperature
\[\langle \mu\rangle =
\Phi\left(\frac{a}{\Lambda},\frac{L}{\Lambda}\right)\]
At $T_C$ the correlation lenght $\Lambda$ goes large, the transition
being 2nd order for $SU(2)$, weak first order for $SU(3)$
\[ \Lambda\mathop\sim_{\tau\to 0} \tau^{-\nu}\]
so that $a/\Lambda \simeq 0$ and $\langle \mu\rangle =
\Phi\left(0,\frac{L}{\Lambda}\right)$. The scaling law follows
\[ \rho/ L^{1/\nu} = f (\tau L^{1/\nu})\]
Using data from different sizes of the lattice $\nu$, $T_C$ and $\delta$
can be determined. 
The result is
\[
\begin{array}{lll}
SU(2) & \nu = .62(1) & \delta = .7(2)\\
SU(3) & \nu = 0.33 & \delta = .5(1)
\end{array}
\]
$\nu$ and $T_C$ agree with other determinations.
In particular it is confirmed that the transition is weak first order for
$SU(3)$, second order for $SU(2)$, in the universality class of the 3d
Ising model.

This result implies that vacuum is superconducting in all the abelian
projected $U(1)$'s.

A typical behaviour of $\mu$ and $\rho$ are shown in fig.1,2. 

\begin{figure}[htb]
\begin{center}
\epsfig{figure=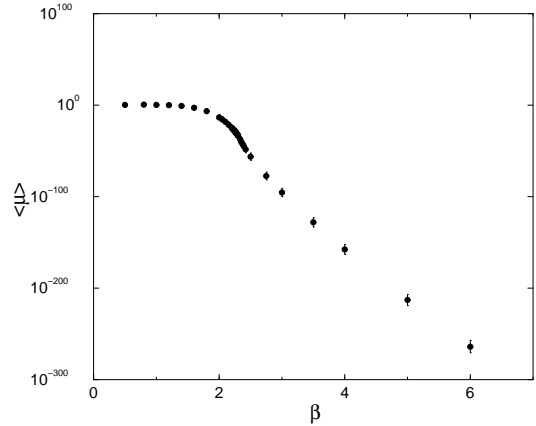,  width=7cm}
\end{center}
\caption{$\langle \mu \rangle$ vs. $\beta$ for $SU(2)$ gauge theory.
Plaquette projection, lattice $16^3 \times 4$.}
\label{fig1}
\end{figure}

\begin{figure}[h]
\begin{center}
\epsfig{figure=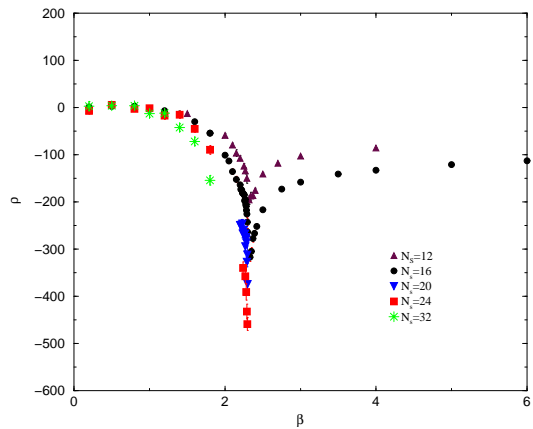, angle=270, width=7cm}
\end{center}
\caption{$\rho$ as a function of $\beta$ for different spatial
sizes.}
\label{rhopla16.fig} \null\vskip 0.3cm
\end{figure}

\section{Vortices.} In 2+1 dimension vortices are local operators
carrying a conserved quantum number\cite{1}. In 3+1 dimension they are
closed defect lines associated to a path $c$. If $B(C)$ is the operator
which creates a vortex on the spatial contour $C$ at time $t$ and $W(C')$
creates a Wilson loop on the contour $C'$ at time $t$, then\cite{1}
\begin{equation}
W(C') B(C) = B(C) W(C') \exp(\frac{2\pi}{N}i n_{CC'})\label{eq6}
\end{equation}
where $ n_{CC'}$ is the winding number of the two paths. Eq.(\ref{eq6})
implies that if $\langle W(C')\rangle$ obeys the area law, i.e. if it
behaves as $\exp(- A_{C'})$ when the size of the path goes large with
respect the correlation length, then $\langle B(C)\rangle$ obeys the
perimeter law, and viceversa if $\langle B(C)\rangle$ obeys the area law
$\langle W(C')\rangle$ will obey the perimeter law. The fact that 
$\langle B(C)\rangle\neq 0$ per se does not imply any simmetry breaking. Some
groups\cite{13} have checked this behaviour by looking at a series of
loops. A systematic procedure\cite{} is to look at the ``dual Polyakov
line'' $\langle \bar L\rangle$, i.e. the stright path wrapping around the
lattice by periodic boundary conditions\cite{14,15}. 
If $\langle
B\rangle$ obeys the area law then $\langle \bar L\rangle = 0$, $\langle
\bar L\rangle \neq 0$ signal perimeter law.

Numerical simulations show that $\langle \bar L\rangle$ is a good order
parameter for confinement, being nonzero in the confined phase, and zero 
in the deconfined one.

Moreover the critical index by which it vanishes with $\tau = 1-t/T_C$,
$\langle \bar L\rangle \sim \tau^\delta$ is $\delta = 0.5\pm.15$, equal within
errors to that of $\langle\mu\rangle$ the disorder parameter
for dual superconductivity\cite{14,15}.
\section{Discussion}
The question: who is $\vec \Phi$? was asked already in ref.\cite{6}.
For long time a zoologically minded answer was given to it.
The relevant abelian projection was taken to be
maximal abelian gauge
because of monopole dominance in that projection.
Investigation
of simmetry, however, shows that all abelian projections are physically
equivalent\cite{10,11}, a possibility already suggested in ref.\cite{6}.

The operator $\mu$ detecting dual superconductivity is defined at the
level of rigor of constructive field theory in compact $U(1)$. In that
case everything is a theorem: numerical simulations were made only to
explore the numerical viability of the method\cite{16}. 
Incidentally $\mu$ is constructed as a gauge invariant charged operator
\`a la Dirac\cite{7}, non local, but local enough to obey the cluster
property, so that
$\langle \mu\rangle\neq 0$ does not violate by any means Elitzur's
theorem\cite{18}.

In non abelian
theories charged fields are present, and in principle the abelian
projection is defined up to terms ${\cal O}(a^2)$, with $a$ the lattice
spacing. This can create a problem at short distances with the Dirac
quantization condition. Looking for a way out can probably suggest
good ideas to identify the dual excitations\cite{19}.

Our conclusion is that dual superconductivity is the mechanism of
confinement. The disorder parameter $\langle\mu\rangle$ detects it, both
in pure gauge and in full QCD\cite{20}, 
as expected in the philosophy of $N_C\to\infty$: indeed quark loops are
non leading in that expansion, and the mechanism of confinement is
expected to be the same with and without quarks.
A carefull analysis of the relation between 
$\langle\mu\rangle$ and the chiral order parameter is on the way.

More investigation is needed to fully understand the dual description of
QCD.

\end{document}